# Isomorphism of graphs- a polynomial test

Moshe Schwartz

School of Physics and Astronomy

Tel Aviv University , Ramat Aviv , 69978 , Israel

An explicit algorithm is presented for testing whether two non-directed graphs are isomorphic or not. It is shown that for a graph of $n$ vertices, the number of $n$ independent operations needed for the test is polynomial in $n$. A proof that the algorithm actually performs the test is presented.

Introduction-The question whether an isomorphism test for two graphs may be found, which is polynomial in the number of vertices, $n$, stands open for quite a while now. The purpose of the present article is to answer this question affirmatively by presenting an algorithm which decides whether two graphs are isomorphic or not and showing that the number of $n$ independent elementary operations needed for that decision is bounded from above by $o(n^6 (\log_2^n)^2)$. The main ingredient in the construction of the algorithm is a theorem stating that if two real $n \times n$ symmetric matrices are such that when raised to any power $k = 1, \ldots, n$ their diagonal elements are identical, the two matrices are identical. This is proven in section 1. In section 2 it is shown how this theorem can be applied to the problem of isomorphism of graphs. This is done essentially by considering the connectivity matrices and asking whether by a properly defined rearrangement of both matrices they can be brought to obey the conditions of theorem I. Section 3 is devoted to the description of an algorithm implementing the ideas presented in section II and to showing that the number of $n$ independent operations needed for the algorithm to answer the question of isomorphism of two graphs is polynomial in $n$.

1. Theorem I-a theorem on real symmetric matrices

Let $\mathbf{A}$ be an $n \times n$ matrix which is: real, symmetric and non-negative ($\mathbf{A}_{ij} \geq 0$). Let $\mathbf{B}$ have the same properties and let $\mathbf{A}$ and $\mathbf{B}$ be related by the conditions $(\mathbf{A}^j)_{ii} = (\mathbf{B}^j)_{ii}$ for all $i$ and $j = 1, \ldots, n$. This implies that $\mathbf{A} = \mathbf{B}$.

Proof:
1.1 The two matrices have the same eigenvalues

Proof: trivial. It is easy to show that all the coefficients of the characteristic polynomial of a matrix $\mathbf{A}$ can be obtained from the quantities: $g_k = tr(\mathbf{A}^k)$ with $k = 1,\ldots,n$. Since the $g_k$'s are identical for both matrices their eigenvalues are the same.

1.2 Let $\lambda_1,\ldots,\lambda_l$ be the set of eigenvalues of $\mathbf{A}$ and $m_1,\ldots,m_l$ the corresponding multiplicities. Let $\mathbf{D}$ be an $n \times n$ diagonal matrix with $m_1$ $\lambda_1$'s, $m_2$ $\lambda_2$'s etc. on the diagonal, ordered in increasing order. It follows from 1.1 that there exist two orthogonal matrices $\mathbf{R}$ and $\mathbf{S}$ such that

$$\mathbf{A} = \mathbf{RDR}^{-1} \text{ and } \mathbf{B} = \mathbf{SDS}^{-1}. \tag{1}$$

Let the first rows of $\mathbf{R}$ and $\mathbf{S}$ be denoted by $\mathbf{r}$ and $\mathbf{s}$ respectively. Let $\mathbf{r}_i$ and $\mathbf{s}_i$ be the projections of $\mathbf{r}$ and $\mathbf{s}$ respectively on the subspace corresponding to the eigenvalue $\lambda_i$ ($\mathbf{r}_1$ consist of the first $m_1$ entries of $\mathbf{r}$, $\mathbf{r}_2$ of the next $m_2$ entries etc.).

The relations $(\mathbf{A}^j)_{11} = (\mathbf{B}^j)_{11}$ imply relations between the two orthogonal matrices

$$\sum_{i=1}^{l} \lambda_i^j x_i = 0, \tag{2}$$

where

$$x_i = \mathbf{r}_i \cdot \mathbf{r}_i - \mathbf{s}_i \cdot \mathbf{s}_i. \tag{3}$$

$$x_i = 0 \text{ for all } i, \tag{4}$$

for which $\lambda_i \neq 0$. From the fact that $\sum_{i=1}^{l} \mathbf{r}_i \cdot \mathbf{r}_i = 1$ follows that $x_i = 0$ also if $\lambda_i = 0$.

The above proof is not limited to the first row. Therefore the conclusion is that if we designate by $\mathbf{r}_i^k$ and $\mathbf{s}_i^k$ the projections of the $k$'s row on the subspace corresponding to $\lambda_i$, then

$$\mathbf{r}_i^k \cdot \mathbf{r}_i^k = \mathbf{s}_i^k \cdot \mathbf{s}_i^k. \tag{5}$$

1.3 The rows of $\mathbf{R}$ (and $\mathbf{S}$) are linearly independent, the set of $n$ $m_i$ dimensional vectors $\{\mathbf{r}_i^k\}$ is linearly dependent, of course, but the dimension of the subspace spanned by that set is $m_i$. (The proof is trivial. Suppose this is not true and the dimension of $\{\mathbf{r}_i^k\}$ is $\alpha < m_i$. This implies that the dimension of $\mathbf{R}$ that is less or equal to $n - m_i + \alpha$, which is a contradiction.)

Therefore, it follows from (5)

$$\mathbf{s}_i^k = \sigma_k \mathbf{r}_i^k \mathbf{T}_i,  \tag{6}$$

where $\mathbf{T}_i$ is an $m_i \times m_i$ orthogonal matrix and $\sigma_k$ may be either + or -1. This implies the orthogonal $n \times n$ matrix, $\mathbf{T}$, that the matrices $\mathbf{R}$ and $\mathbf{S}$ are related by

$$\mathbf{S} = \sigma \mathbf{R} \mathbf{T},  \tag{7}$$

where the orthogonal $n \times n$ matrix, $\mathbf{T}$, is a block matrix with the $\mathbf{T}_i$'s as blocks, arranged along the diagonal in increasing order of $i$ and $\sigma$ is a general diagonal matrix with $\pm 1$ on the diagonal.

Let $\Omega_i$ be the eigen-subspace of $\mathbf{A}$ corresponding to the eigenvalue $\lambda_i$ (the space of column vectors $\{\mathbf{v}_\alpha^i\}$ obeying $\mathbf{A}\mathbf{v}_\alpha^i = \lambda_i \mathbf{v}_\alpha^i$). Let $B_i$ be a **general** orthonormal basis for $\Omega_i$. The most general form of the matrix $\mathbf{R}$ is obtained by arranging from left to right the $m_1$ vectors of $B_1$ then the $m_2$ vectors of $B_2$ etc. Let $\mathbf{R}_1$ and $\mathbf{R}_2$ be two such matrices then, clearly a block orthogonal matrix $\overline{\mathbf{T}}$ can be found with $m_i \times m_i$ blocks along the diagonal arranged in increasing order of $i$, such that

$$\mathbf{R}_2 = \mathbf{R}_1 \overline{\mathbf{T}}.  \tag{8}$$

It follows from equation (7) that the matrix $\sigma \mathbf{S}$ is exactly of the general form of $\mathbf{R}$ described above.

Therefore

$$\mathbf{B} = \mathbf{S}\mathbf{D}\mathbf{S}^{-1} = \sigma \mathbf{R}\mathbf{D}\mathbf{R}^{-1}\sigma = \sigma \mathbf{A} \sigma.  \tag{9}$$

Since $\mathbf{A}$ and $\mathbf{B}$ are both non-negative it follows at once that all the diagonal elements of $\sigma$ must have the same sign and therefore

$$\mathbf{A} = \mathbf{B}.  \tag{10}$$

2. Theorem II- application to the problem of isomorphism of non-directed graphs

2.1 Definition: Self connectivity of order $k$ of a given vertex of a non-directed graph $G$ is the number of paths of $k$ steps connecting the vertex to itself. Denote the self connectivity of order $k$ of the vertex $i$ by $N_i^k$.

2.2 Elementary result: Let $\mathbf{A}$ be the connectivity matrix of the graph $G$ of $n$ vertices (A symmetric matrix with one in the entry $ij$ if the vertices $i$ and $j$ are connected by an edge in $G$ and zero if they are not connected) then

$$N_i^k = (\mathbf{A}^k)_{ii}.  \tag{11}$$

2.3 Definition: A non-directed graph, $G_k$, is a $k$ rearrangement of the non-directed graph $G$ if it is obtained from $G$ by re-labeling the vertices so that in $G_k$ $i < j$ implies one of the two following options: (1) There exists $l < k$ such that $N_i^m = N_j^m$ for all $m < l$ and $N_i^l < N_j^l$. (2) The self connectivities are equal for all orders less than $k$ ( $N_i^m = N_j^m$ for all $m < k$) and $N_i^k \leq N_j^k$. ( Clearly, given a graph $G$, $G_k$ is not necessarily unique for any $k$.)

2.4 Definition: A $k$ diagonal of $G$, $D(G_k)$, is a matrix defined by

$$D(G_k)_{ij} = N_i^k \delta_{ij}, \text{ where} \tag{12}$$

the $N_i^k$'s correspond to some $k$ rearrangement of $G$, $G_k$.

2.5 Elementary result: Let $G_k^{(1)}$ and $G_k^{(2)}$ be two $k$ rearrangements of the graph $G$ then $D(G_m^{(1)}) = D(G_m^{(2)})$ for $m \leq k$. Therefore for $m \leq k$ $D(G_m)$ can be written as $D_m(G)$ to denote that it depends only on $G$.

2.6 Elementary result: Let $\overline{G}$ and $\tilde{G}$ be two isomorphic graphs. Since isomorphism implies that one is obtained from the other by re-labeling of the vertices it is clear that $D_k(\overline{G}) = D_k(\tilde{G})$ for all $k$.

2.7 Theorem II: Let $\overline{G}$ and $\tilde{G}$ be two graphs such that $D_k(\overline{G}) = D_k(\tilde{G})$ for all $k \leq n$ then the two graphs are isomorphic.

   Proof: From equations (11) and (12) it follows that the matrices $\overline{\mathbf{A}}_n$ and $\tilde{\mathbf{A}}_n$ corresponding to $n$ rearrangements of $\overline{G}$ and $\tilde{G}$ respectively are symmetric, obey that $((\overline{\mathbf{A}}_n)^j)_{ii} = ((\tilde{\mathbf{A}}_n)^j)_{ii}$ for all $i, j = 1, \ldots, n$ and are obviously non-negative. By Theorem I, $\overline{\mathbf{A}}_n = \tilde{\mathbf{A}}_n$. Since $\overline{G}_n$ is isomorphic to $\overline{G}$ and $\tilde{G}_n$ is isomorphic to $\tilde{G}$, the proof that $\overline{G}$ is isomorphic to $\tilde{G}$ is complete.

3. Algorithm for testing whether two non-directed graphs are isomorphic
 In the following I describe an algorithm for checking whether two graphs of $n$ vertices are isomorphic. The algorithm to be presented is based on the material discussed above. It is not the most efficient one that can be constructed on that basis but easy to describe and to prove that it is bounded from above by a power law in $n$.

3.1(a)Obtain the matrices $\overline{\mathbf{A}}^2$ and $\tilde{\mathbf{A}}^2$ where $\overline{\mathbf{A}}$ and $\tilde{\mathbf{A}}$ are the connectivity matrices of the two graphs. Re-label the vertices to order that the diagonal elements of both matrices $\overline{\mathbf{A}}^2$ and $\tilde{\mathbf{A}}^2$ in increasing order. This yields the matrices $\mathbf{D}_2(\overline{G})$ and $\mathbf{D}_2(\tilde{G})$. If the two matrices are not identical, stop. The two graphs are not isomorphic. If they are identical we have along the diagonals $\alpha$ different values of the second order self connectivity arranged in increasing order with multiplicities $m_2^1, \ldots, m_2^\alpha$. If all the multiplicities are 1, stop. The two graphs are isomorphic. (c) If the **D** 's are identical but not all the multiplicities are one construct the two matrices corresponding to the two original graphs after the re-labeling of the vertices, $\overline{\mathbf{A}}_2$ and $\widetilde{\mathbf{A}}_2$. (d) The last two matrices are raised to the third power and the vertices are re-labeled in the following way. The first $m_2^1$ vertices are re-labeled in increasing order according to their third order self connectivities, then he next $m_2^2$ are re-labeled etc. This yields the matrices $\mathbf{D}_3(\overline{G})$ and $\mathbf{D}_3(\widetilde{G})$. (e) If the two matrices are not identical, stop. The two graphs are not isomorphic. If they are identical we have along the first $m_2^1$ entries of the diagonal $\beta 1$ different values of the third order self connectivity with multiplicities $m_3^1, \ldots, m_3^{\beta_1}$. Along the next $m_2^2$ entries of the diagonal we have $\beta 2$ different values of the third order self connectivity with multiplicities $m_3^{\beta_1+1}, \ldots, m_3^{\beta_1+\beta_2}$ etc.. If all the $m$ 's are 1, stop. The two graphs are isomorphic. If the two $\mathbf{D}_3$ 's are identical but not all the $m_3$ 's are one, construct the two matrices corresponding to the original graphs after the last re-labeling of the vertices $\overline{\mathbf{A}}_3$ and $\widetilde{\mathbf{A}}_3$. (f) These are raised now to the fourth power etc. (g) A decision , if not reached before that, must be reached after obtaining $\mathbf{D}_n(\overline{G})$ and $\mathbf{D}_n(\widetilde{G})$. If the two $\mathbf{D}_n$ 's are identical the graphs are isomorphic if not they are not.

3.2 A bound on the number of elementary operations
The algorithm involves four basic steps: (a) Raising an $n \times n$ matrix to a power of up to $n$. The number of operation involved is bounded by $o(n^4)$. (b) Rearranging sets of up to $n$ numbers up to $n$ times (arranging the **D** 's). The number of operations is bounded by $o(n^3)$. (c) Rearranging of rows and columns. (This is done to obtain the $\mathbf{A}_k$ 's). This may be done up to $n$ times. The number of operations is bounded by $o(n^3)$. (d) Checking if the **D** 's are identical. This involves comparing up to $n$ numbers up to $n$ times. The number of operations is bounded by $o(n^3)$. Consequently the total number of elementary operations is bounded by $o(n^4)$. Since the numbers involved in the various operations may become rather large with $n$, each elementary operation with those number may have in itself a strong dependence on $n$. Thus the relevant $n$ dependence must include also the dependence of an elementary operation, such as multiplication of two big , $n$ dependent numbers. Since the numbers involved as entries of the matrices are bounded by $n^n$, it is not difficult to realize that the total number of $n$ independent operations is bounded by $(n\log_2^n)^2 \times$ bound on the total number of operations. Thus the total number of $n$ independent operations is bounded by $o(n^6(\log_2^n)^2)$. This is rather crude but enough to prove the statement that it is polynomial in $n$.